\documentclass[%
 reprint,
 amsmath,amssymb,
 aps,
]{revtex4-2}

\usepackage{graphicx}% Include figure files
\usepackage{dcolumn}% Align table columns on decimal point
\usepackage{bm}% bold math
\usepackage[capitalise]{cleveref}
\UseRawInputEncoding

\begin{document}
\preprint{APS/123-QED}

\title{Evolution of the number and temperature of the remaining cold atoms in CW-laser photoionization of laser-cooled $^{87}$Rb atoms}% Force line breaks with \\
%\thanks{A footnote to the article title}%

\author{Fei~Wang$^{1,a}$, Feng-Dong~Jia$^{2,a}$}
\thanks{$^a$These authors have contributed equally to this work.}

\author{Wei-Chen~Liang$^2$}

\author{Xiao-Kang~Li$^2$}

\author{Yu-Han~Wang$^2$}

\author{Jing-Yu~Qian$^2$}

\author{Dian-Cheng~Zhang$^3$}

\author{Yong~Wu$^4$}

\author{Jian-Guo~Wang$^4$}

\author{Rong-Hua~Lu$^5$}

\author{Xiang-Yuan~Xu$^6$}

\author{Ya-Ping~Ruan$^7$}
\email{ruanyaping@nju.edu.cn}

\author{Ping~Xue$^1$}
\email{xuep@tsinghua.edu.cn}

\author{Zhi-Ping~Zhong$^2$}
\email{zpzhong@ucas.ac.cn}

\affiliation{$^1$State Key Laboratory of Low-Dimensional Quantum Physics, Department of Physics, Tsinghua University, Beijing 100084,China}
\affiliation{$^2$School of Physical Sciences, University of Chinese Academy of Sciences, Beijing 100049, China}
\affiliation{$^3$Department of physics, Beijing Normal University, Beijing 100875, China}
\affiliation{$^4$Institute of Applied Physics and Computational Mathematics, Beijing 100088, China}
\affiliation{$^5$Shanghai Institute of Optics and Fine Mechanics (SIOM), Chinese Academy of Sciences (CAS), Shanghai 201800, China}
\affiliation{$^6$Department of Physics, Capital Normal University, Beijing 100037, China}
\affiliation{$^7$National Laboratory of Solid State Microstructures, College of Engineering and Applied Sciences, Nanjing University, Nanjing 210093, China}
\date{\today}% It is always \today, today,
 % but any date may be explicitly specified

\begin{abstract}

Based on the Rb$^+$-Rb hybrid trap, we investigate the effect of ion-atom elastic collisions on the number and temperature of the remaining atoms. We measured the remaining atomic number and temperature as a function of the wavelength and intensity of the ionization laser, and whether the ion trap was turned on. Fittings with a single exponential decay function plus an offset to the number and radius of the remaining atoms are found to be in good agreement. We found a difference in the exponential factor of different wavelengths of ionization laser with the ion trap on or off. We suppose that the presence of electrons affects ion-atom collisions through disorder-induced heating. Our research contributes to a better understanding of how ultracold neutral plasma evolves, particularly the subsequent kinetics of atomic processes, which also serves as a useful reference for high-energy-density plasma.

\begin{description}
%\item[Usage]
%Secondary publications and information retrieval purposes.
\item[PACS numbers]
34.50.Cx,34.80.Dp,34.90.+q,52.20.Hv,52.27.Gr
%May be entered using the \verb+\pacs{#1}+ command.
%\item[Structure]
%You may use the \texttt{description} environment to structure your abstract;
%use the optional argument of the \verb+\item+ command to give the category of each item.
\end{description}
\end{abstract}

\pacs{Valid PACS appear here}% PACS, the Physics and Astronomy
 % Classification Scheme.
%\keywords{Suggested keywords}%Use showkeys class option if keyword
    %display desired
\maketitle
%\tableofcontents

\section{Introduction}

Photoionization is an important process in many fields of science, such as atomic and molecular physics, astrophysics, plasma physics, and atmospheric science. The mixture of electrons, ions, and neutral particles created by photoionizing atoms or neutral molecules provides a powerful tool for understanding the structure and dynamics of complex physical systems, e.g., the collisions in the mixture. Laser-cooled atoms offer great opportunities for studies of precise spectroscopic measurements, quantum coherent phenomena, and low-energy collisions with the neutral atoms due to the significant mitigating of the Doppler broadening effect, and low collision rate. 

Atom-trap-based techniques have been widely used to measure the absolute cross-section for collisional processes and photoionization processes in a magneto-optical trap (MOT), such as pioneering work in electron collisional processes \cite{motcross1995,motcross1996} and photoionization process\cite{motIP1992}. This is because collisions typically involve a change in the kinetic energies, velocities, and/or chemical structure of the collision partners, as well as trap loss; a photoionization process also results in trap loss. Ruan \emph{et al.}\cite{ruan-2014} discovered cold ion-atom collisions heat the remaining atoms by extending the trap-loss measurement to measure both the temperature and number of remaining atoms in the two-step CW-laser photoionization of a laser-cooled $^{87}$Rb cloud in a standard vapor-loaded MOT with a glass chamber. Furthermore, ultracold-neutral plasmas (UNPs) can be created by pulsed photoionization of laser-cooled atoms near the ionization threshold\cite{UNP1999}. UNP is an effective high-energy-density plasma (HEDP) simulator because they overlap in the strongly-coupled region and have similar coupling parameter $\Gamma$ and screening length $\kappa$ \cite{UNP2007,UNP2017,bergeson_2019}. Collisions are crucial in the evolution of a UNP. However, up to now, only electron-ion collisions are considered in the study of ultracold neutral plasmas, and interactions of the charged particles with the neutral atoms are neglected since Killian \emph{et al.}\cite{UNP1999} believe that the mean free path for neutral-charged particle collisions is much larger than the sample size (typically on the order of 0.5 mm)\cite{UNP1999}. This was verified in electron-atom collisions based on the calculations for elastic scattering of slow electrons from noble gases ($\sim 38$ a.u. at 1 K)\cite{eleccross1988}; however, ion-neutral collisions dominated by long-range interactions between atoms and ions by a factor of $C_4/r^{-4}$ \cite{RMP2019,heazlewood-2021,dieterle_transport_2021,puri_reaction_2019}. Here $C_4$ is the leading long-range induction coefficient and $r$ is the internuclear distance. As a result, large elastic scattering cross-sections ($\sim 10^6$ a.u. at 1 mK)\cite{hyb2014,RMP2019} are expected, allowing for strong ion-atom interactions.

Owing to the advantages of laser trapping, cooling, and ion-trapping techniques, cold hybrid ion-atom systems have emerged over the past 20 years, paving the way for the study of ion-atom collisions in the quantum regime. The hybrid system provides highly controllable quantum systems with tunable ion-atom long-range interactions, and the theoretical and experimental progress has been well summarized by the most recent review\cite{RMP2019}. Due to the ion-atom large elastic scattering cross-section, a wide range of exciting experiments have been proposed, such as reaching ultra-low temperatures with sympathetic cooling, ultracold charge transport, new many-body bound states, and strongly coupled polaritons, quantum information processing, and quantum simulations, etc.\cite{krukow_reactive_2016,dieterle_transport_2021,puri_reaction_2019} The majority of these studies rely on interactions mediated by elastic ion-atom collisions, but little is known about the subsequent kinetics of ion-atom collision processes\cite{mohammadi_life_2021}.

In the two-step CW-laser photoionization, these ion-atom collisions might be reactive, inelastic, or elastic. In this study, we investigate how the elastic processes develop and influence the number and temperature of the remaining atoms. The present work is an in-depth discussion of Ruan \emph{et al.}\cite{ruan-2014}. Using atomic absorption imaging techniques and adjusting the system parameters, we measured the change in the number and temperature of the remaining atoms as a function of system parameters during the two-step CW-laser photoionization of the laser-cooled $^{87}$Rb atoms in the ion-neutral hybrid trap. The system parameters include the wavelength and intensity of the ionization light, and whether the ion trap is turned on. These in-depth studies of ion-atom cold collisions contribute to a deep understanding of UNPs’ evolution, the physics of strongly correlated many-body systems, quantum simulations, etc.

\section{Method}

The detailed description of our ion-neutral hybrid trap for rubidium atoms can be found in our previous study\cite{Lv2016cpl}. This trap consists of an Rb standard MOT and a mass-selective linear Paul trap (LPT), which are concentrically arranged in a polyhedral flat non-magnetic stainless-steel cavity. The following steps are performed during each experimental cycle in the ion-atom hybrid trap: the MOT is loaded to a steady state, the CW ionization light is turned on for a predetermined irradiation time either the ion trap is on or off, the CW ionization light is switched off, and the trapped ions are pushed to the MCP by turning off the voltage on the end-cap ring electrode closer to the MCP. The ion time-of-flight (TOF) spectrum is recorded by an oscilloscope. In the meantime, the TOF approach is utilized to obtain the temperature of the cold atoms while atomic absorption imaging techniques are used to count the atoms and measure the radius of atomic clouds.

System parameters are listed below. The atom number and radius $r_{atom}$ (1/$e^2$ half-waist) of the cold atomic cloud were measured as $\sim 5\times 10^7$, $\sim 0.5$ mm by absorption imaging. The first excitation laser was the MOT cooling laser with a detuning of -12 MHz from the resonant frequency of $5~^2S_{1/2},F=2\rightarrow 5~^2P_{3/2},F'=3$ transition. The second excitation laser, that is, the ionization laser, was provided by another CW-diode laser with a variable wavelength in the range of $\lambda_{ion}=$ 450$\sim$479 nm. The radial directions x, y, and axial direction z of the ion cloud were 2.32, 2.32, and 20.20 mm, respectively\cite{Lv2016cpl,XiaoKang2020}. The trap depth of the ion trap is approximately 0.7 eV, corresponding to a maximum temperature ranging at $10^3-10^4$ K for the trapped Rb$^+$ ion.

Due to the need to fulfill the laws of energy and momentum conservation, as well as the fact that the electron mass is much smaller than the ion mass, the initial temperature of the ion is slightly higher than the atomic temperature and is around mK. The majority of the excess photon energy, i.e., the difference between the photon energy and the ionization threshold, is carried by electrons. The wavelength of the ionization laser is varied among 447, 450, 475, 476, 477, 478, 478.8, and 479 nm, corresponding to initial electron temperatures of 1438.5, 1295.2, 171.4, 128.9, 86.6, 44.4, 10.8 and 2.5 K, respectively. In the case of UNPs created by pulsed photoionizing laser-cooled atoms near the ionization threshold, the ions are then heated to several K on a time scale of approximately $\sim 10^2$ ns by disorder-induced heating (DIH)\cite{UNP2007}. DIH is related to the spatial distribution of the charged particles during ionization\cite{UNP2007,UNP2017}. Thus, the ion-atom collision energy $E_{col}$ in our experiment is roughly 10 mK-K. Côté and Dalgarno\cite{Dalgarno-2000} obtained the expression for the elastic cross-section for ultracold atom-ion collisions as $\sigma_{ela.}\propto E_{col}^{-1/3}$. Thus the ion-atom elastic rate constant follows $K=\langle \sigma_{ela.} v\rangle\propto E_{col}^{1/6}$ ($\langle\cdot\rangle$ indicates averaging over velocities) and is proportional to the collision energy, which is nearly equal to ion temperature.

\begin{figure}[htbp]
	\includegraphics[width=3.5in]{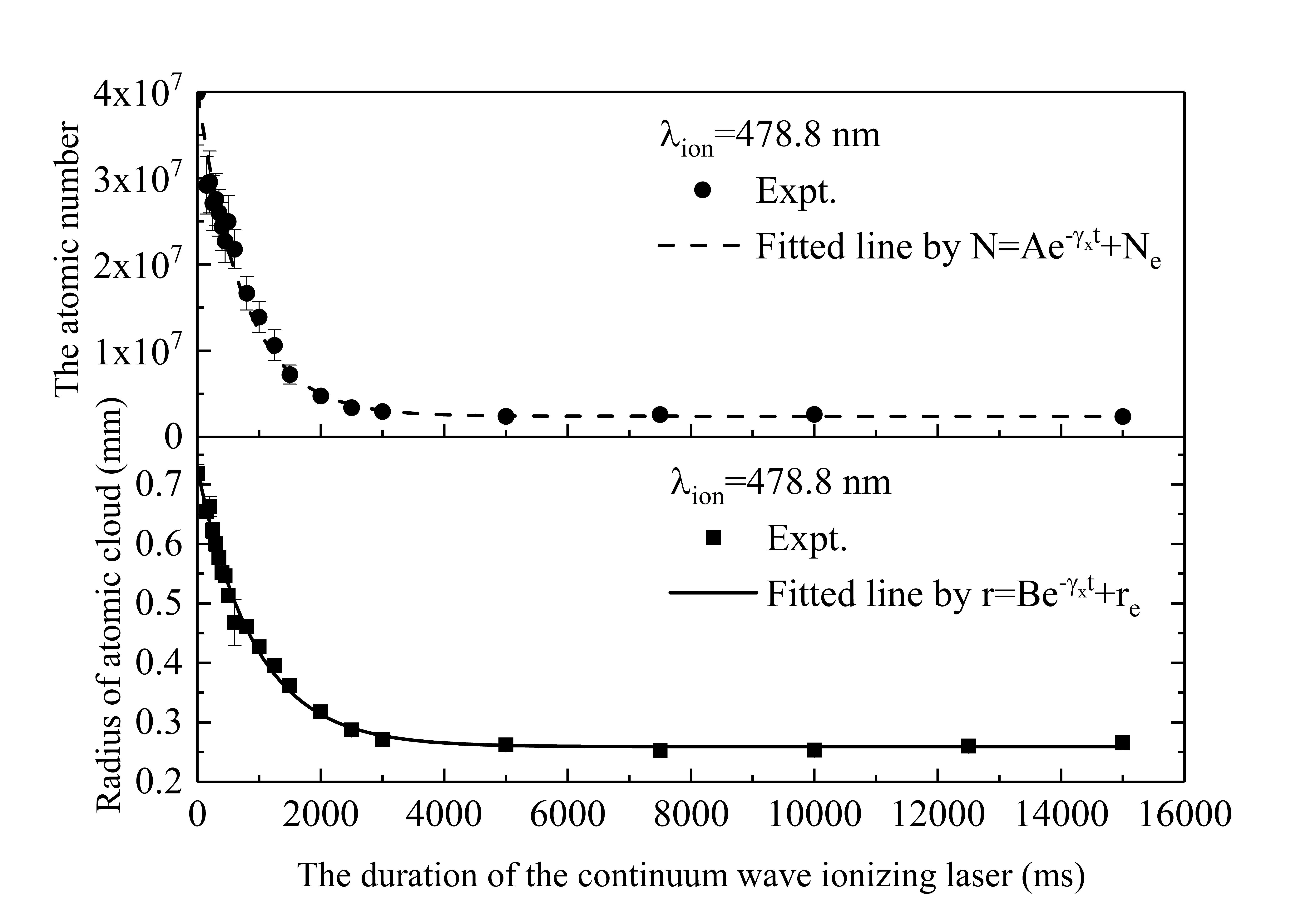}
	\caption{\label{atom-size}The number (up) and the radius (down) of atomic cloud in the MOT as the function of the irradiation of the ionization laser in the two-step CW-laser photoionization of laser-cooled $^{87}$Rb atoms in the ion-neutral hybrid trap \cite{Lv2016cpl}. The wavelength of the ionization laser is 478.8 nm. Time zero is the moment that the ionization laser turns on and the following graphs are so.}
\end{figure}

\begin{figure}[htbp]
	\includegraphics[width=3.5in]{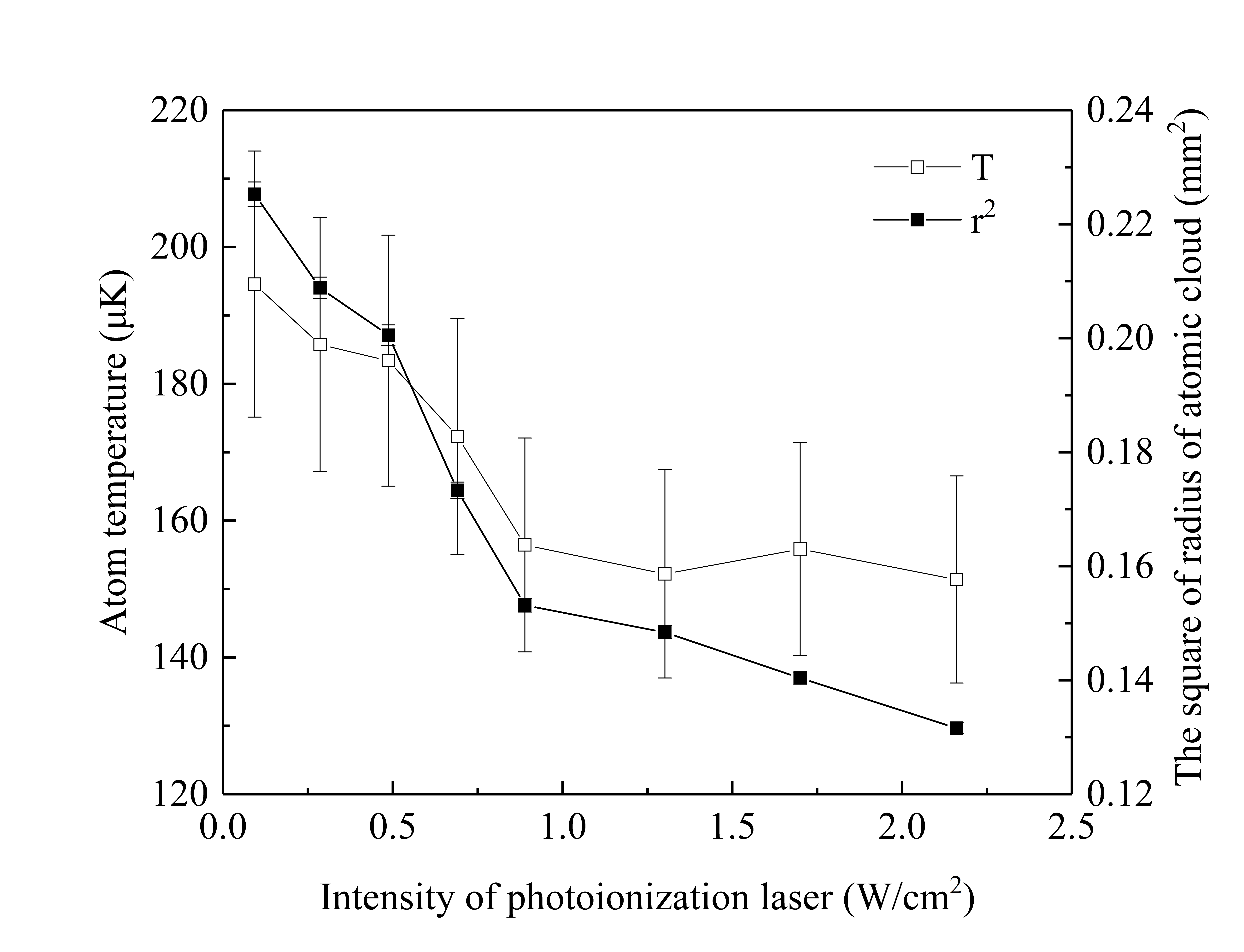}
	\caption{\label{size-T}Comparison of the cold atom temperature measured by the time-of-flight method and the square of the radius of the cold atomic cloud measured by atomic absorption imaging techniques. Measurements are performed in the two-step CW-laser photoionization of laser-cooled $^{87}$Rb atoms in ion-neutral hybrid trap\cite{Lv2016cpl}.}
\end{figure}

\begin{figure}[htbp]
	\includegraphics[width=3.5in]{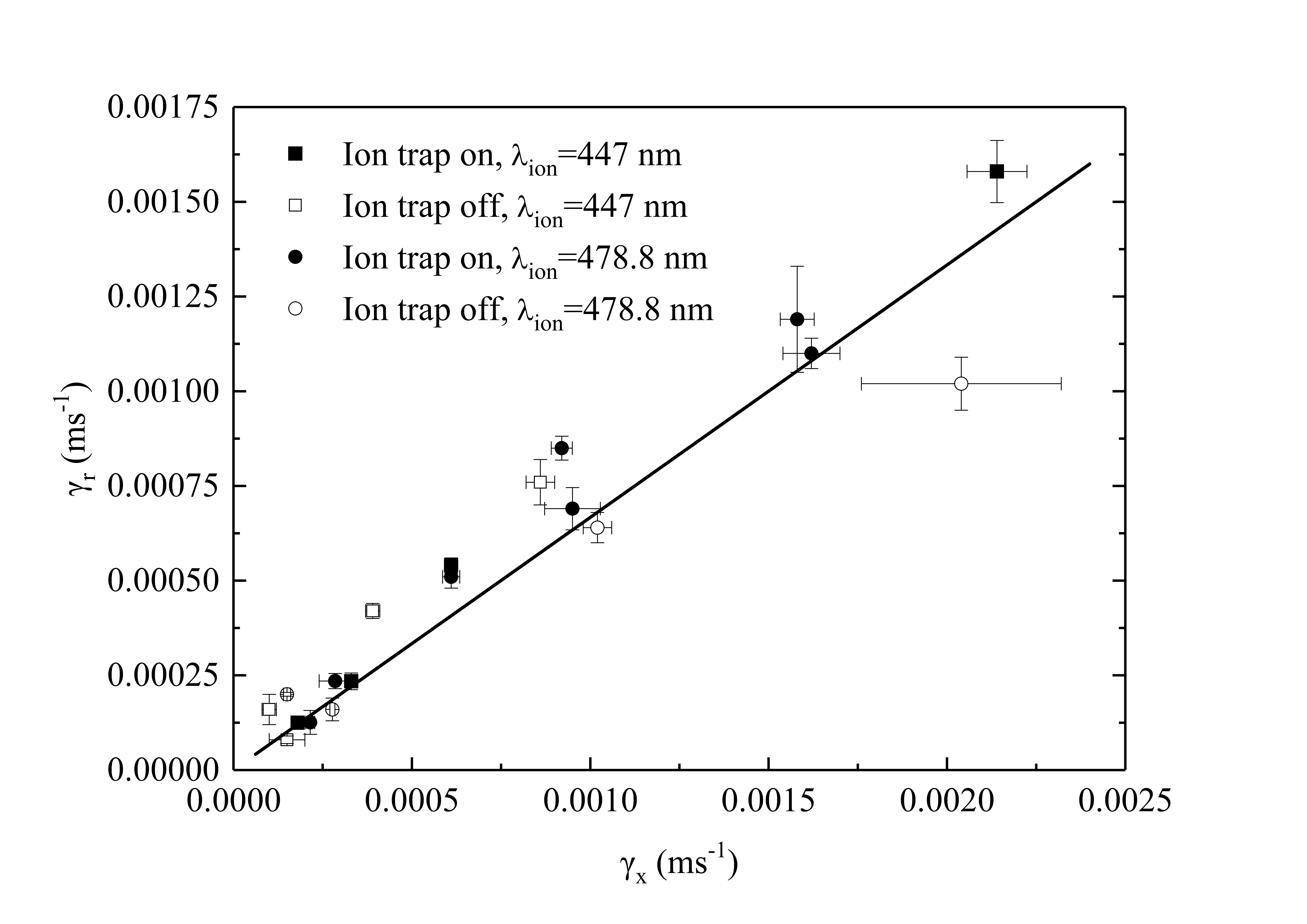}
	\caption{\label{N-T}$\gamma_r$ as a function of $\gamma_x$ measured at different intensities and wavelengths of the ionization laser with the ion trap on or off.}
\end{figure}

The atomic number variation with time can be fitted with a single exponential decay function plus offset according to the rate equation of the total atomic number\cite{Fuso-2000,Rangwala-pra2013,Lv2016cpl}
\begin{equation}\label{eq:atomloss}
	N_{atom}(t)=Ae^{-\gamma_x t}+N_e, \gamma_x=\gamma_L+\gamma_{PI}+\gamma_{ia}.
\end{equation}	

\begin{equation}\label{eq:lossPI}
\gamma_{PI}=\frac{f\sigma_{ion}}{E_{ion}} I_{PI},f=\frac{I/I_s}{1+2I/I_s+(2\delta/\Gamma)^2}.
\end{equation}
Here $I_{PI}$ and $I$ stand for the intensity of the ionization laser and the total cooling laser, respectively, $E_{ion}$ for ionization laser photon energy, $f$ for excited state fraction, and $\sigma_{ion}$ for photoionization cross-section. $I_s$ is the saturation intensity the transition $5~^2S_{1/2},F=2\rightarrow 5~^2P_{3/2},F^\prime=3$. $\delta$ is the detuning of the cooling laser frequency from resonances, which is -12MHz in the present experiment. The offset $N_e$ represents the equilibrium number of atoms, and the exponential factor $\gamma_x$ is the loss rate of the MOT atoms, including $\gamma_L$ caused by collisions between cold atoms, $\gamma_{PI}$ caused by photoionization of cold atoms, and $\gamma_{ia}$ caused by ion-atom collisions. It's interesting to note that this kind of analytical expression can also describe radius $r_{atom}$ of the cold atomic cloud, as shown in Fig.\ref{atom-size}.
\begin{equation}\label{eq:atomsize}
	r_{atom}=B e^{-\gamma_r t}+r_e.	
\end{equation}
Here the offset $r_e$ represents the equilibrium radius of the cold atomic cloud. The exponential factor $\gamma_r$ is the reduction rate of atomic cloud radius and means the decreasing rates of the temperature of remaining atoms. The relation between atomic temperature and the radius of an atomic cloud can be expressed as $T\propto r_{atom}^2$ \cite{book1999}, as illustrated in Fig.\ref{size-T}. Furthermore, our experimental results show that the exponential factors $\gamma_x$ and $\gamma_r$ are well linearly related when measured at various ionization laser intensities and wavelengths with the ion trap on or off, as shown in Fig.\ref{N-T}. This is because, whereas the number of atoms in a magneto-optical trap and atomic temperature generally follow a power law \cite{n-T-1,n-T-2,n-T-3,n-T-4,n-T-5,n-T-6}.

The experimental error in this study resulted from the following factors: the approximately 10-15\% systematic error from the fluctuation of temperature and number of cold atoms, the approximately 1-5\% error resulting from the deconvoluting procedure, the statistical uncertainties, the uncertainties in determining the intensity of the ionization laser, and the uncertainties in determining the frequency/wavelength of lasers. Specifically, 1 nm for $\lambda_{ion}$ = 447 and 450 nm, 600 MHz for $\lambda_{ion}$ = 475-479 nm.

\section{Result}

We investigate how the ion-atom elastic collision and the presence of electrons affect the number and temperature of the remaining atoms. We first discuss the exponential factors $\gamma_x$ and $\gamma_r$ that inscribe the rate of evolution of the atomic number and temperature, respectively. We performed these experiments as the function of the irradiation period of the ionization laser in the Rb$^+$-Rb hybrid trap, and the wavelength of the ionization laser is 447 nm or 478.8 nm with the ion trap on or off. Next, we survey the variation of atomic number and temperature with the intensity of the ionization laser in the case that the irradiation period of the ionization laser was set as 4 s, with the ion trap on or off, and the wavelength of the ionization laser was varied as 450, 475, 476, 477, 478 and 479 nm, corresponding to initial electron temperatures of 1295.2, 171.4, 128.9, 86.6, 44.4, and 2.5 K, respectively.

\begin{figure}[htbp]
	\includegraphics[width=3.5in]{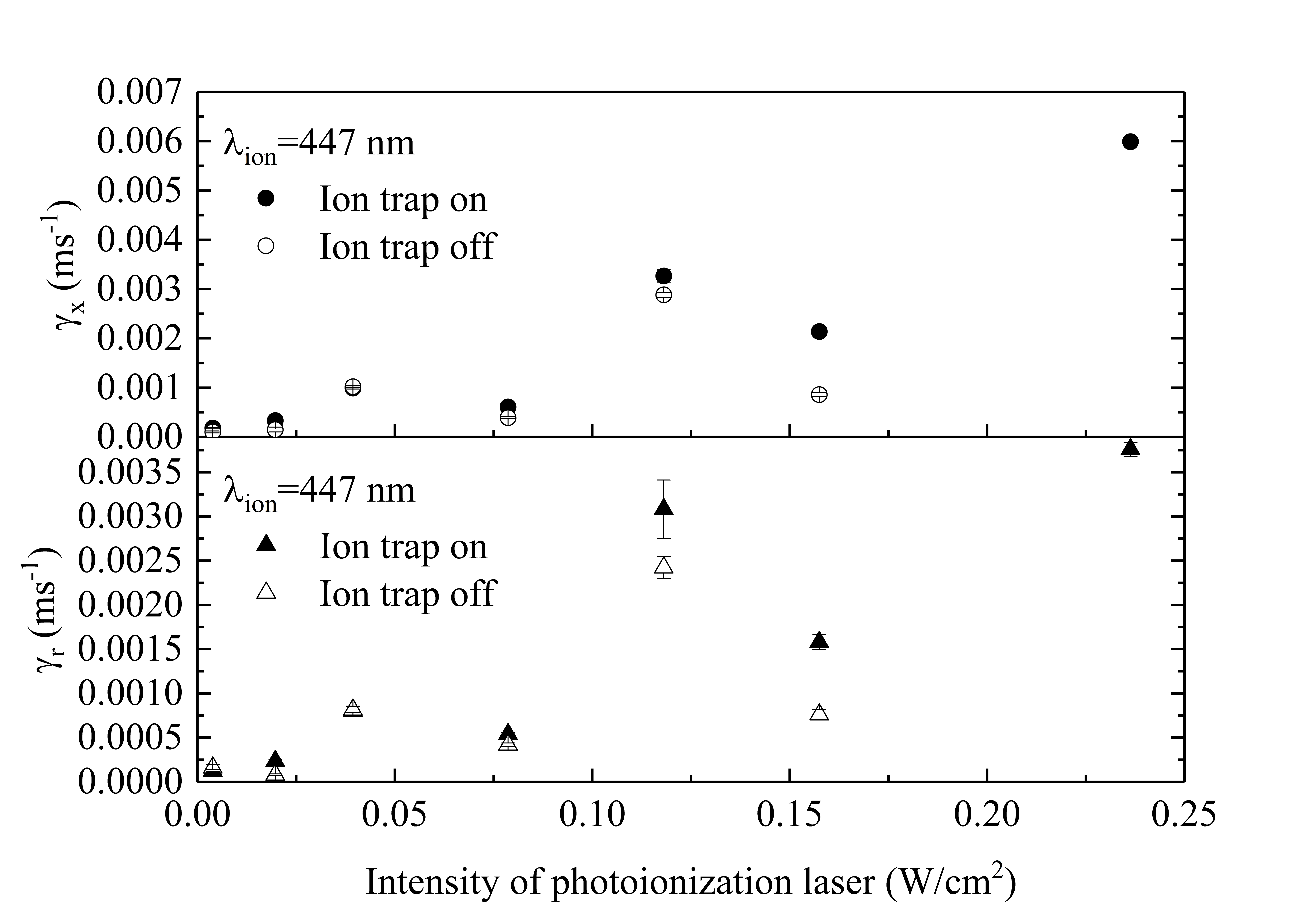}
	\caption{\label{N-T-1A}Comparison of exponential factors $\gamma_x$ and $\gamma_r$ as the function of the ionization laser intensity at $\lambda_{ion}$=447 nm, ion trap on or off.}
\end{figure}

\begin{figure}[htbp]
	\includegraphics[width=3.5in]{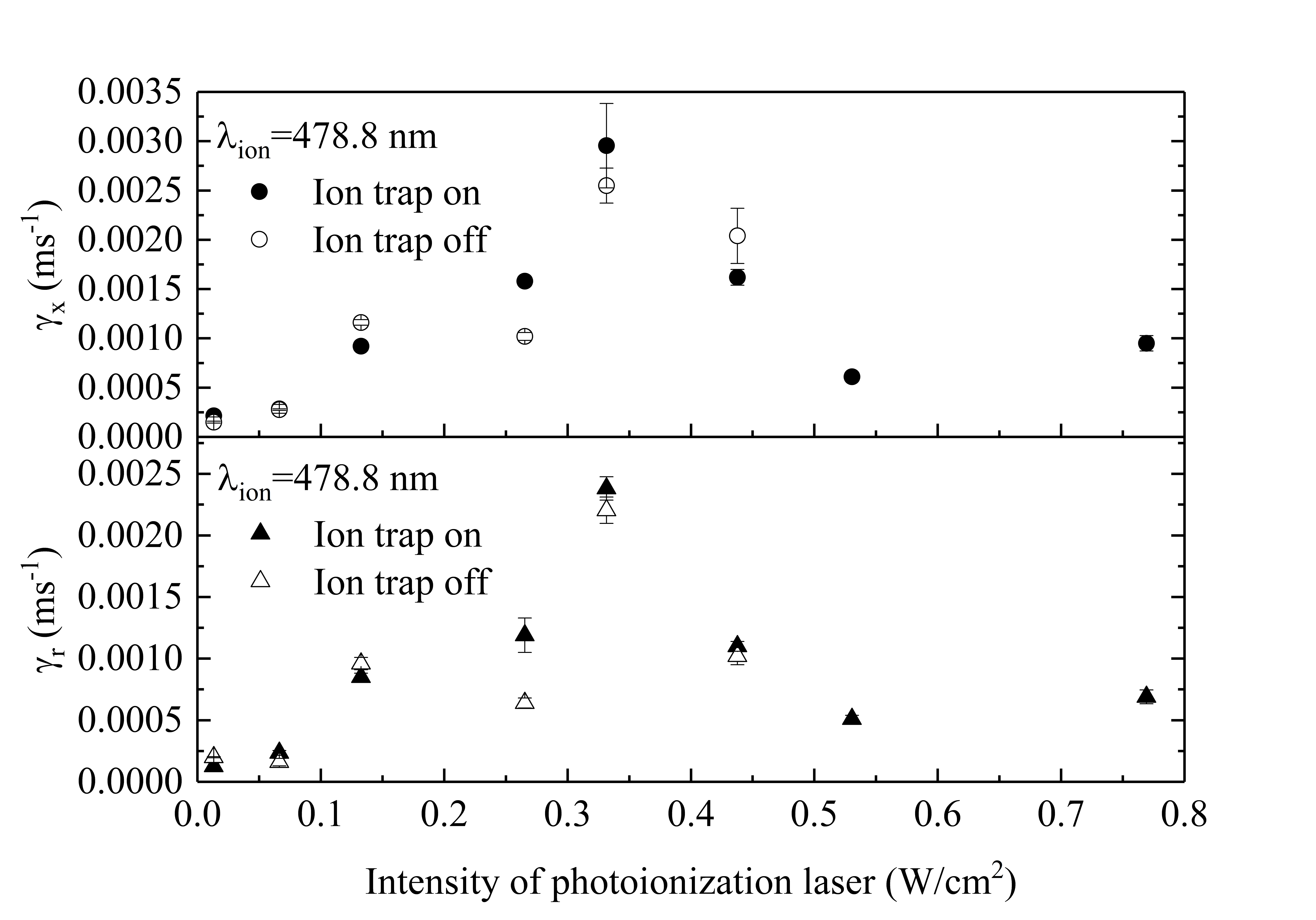}
	\caption{\label{N-T-1B}Comparison of exponential factors $\gamma_x$ and $\gamma_r$ as the function of the ionization laser intensity at $\lambda_{ion}$=478.8 nm, ion trap on or off.}
\end{figure}

As shown in Fig.\ref{N-T-1A}, both two exponential factors $\gamma_x$ and $\gamma_r$ in the ion trap-on case are larger than those in the ion trap-off case with $\lambda_{ion}$ = 447 nm. However, the two exponential factors $\gamma_x$ and $\gamma_r$ when the ion trap is on are not always larger than those in the ion trap off case with $\lambda_{ion}$ = 478.8 nm, as shown in Fig.\ref{N-T-1B}. We will discuss the mechanisms underlying these experimental phenomena as follows. Firstly, the loss rate caused by ion-atom collisions $\gamma_{ia}$ can be divided into $\gamma_{ia}^{MOT}$ and $\gamma_{ia}^{LPT}$, as shown in the following Eq.\ref{eq:atomloss2}.
\begin{equation}\label{eq:atomloss2}
	\gamma _x=\gamma _L+\gamma _{PI}+\gamma _{ia}^{MOT}+\gamma _{ia}^{LPT}.
\end{equation}
$\gamma_{ia}^{MOT}$ is the loss rate of the MOT atoms caused by ion-atom collisions from the ions in the MOT area. $\gamma_{ia}^{LPT}$ is the loss rate of the MOT atoms caused by ion-atom collisions from the ions in the ion trap area. Certainly, $\gamma_{ia}^{LPT}$ is zero when the ion trap is off. Now we discuss how the presence of electrons affects ion-atom elastic collision. The study of ultracold neutral plasma shows that the temperature of the ions is rapidly heated to several K due to disorderly induction heating (DIH). Plasma can only be created when the initial electron temperature is less than 1000 K due to the size of the MOT atom. DIH is brought on by electron-ion spatial correlations \cite{UNP2007,UNP2017}. Therefore, electron-ion spatial correlations cannot be established, and DIH is reduced or even eliminated completely. Thus, regardless of the initial electron temperature, the ion trap repels the electrons, making it difficult to establish electron-ion spatial correlations. Thus, the heating caused by DIH is also reduced or eliminated entirely when the ion trap works. In the case of $\lambda_{ion}$ = 447 $ nm, \gamma_{ia}^{LPT}$  is equal to zero when the ion trap is turned off, and becomes $\gamma_{ia}^{LPT}>0$ when the ion trap is turned on. Therefore, the atomic loss rate $\gamma_x$ in the ion trap-on case is larger than that in the ion trap-off case as shown in Fig.\ref{N-T-1A}. As discussed above, the exponential factors $\gamma_x$ and $\gamma_r$ are well linear, so a similar conclusion is made for $\gamma_r$. As a contrast, when the $\lambda_{ion}=$ 478.8 nm, the initial electron temperature is 10.8 K, our result shows that occasionally greater than those in the ion trap-off, and occasionally smaller. We hypothesize that the ions in the MOT area are heated by DIH in the ion trap-off situation. Since the ion-atom elastic rate constant is proportional to the collision energy as discussed above, when the ion trap turns off, the ion temperature in the MOT is higher than it would be in the case of the ion trap on. As a result, the ion-atom elastic rate constant is higher than it would be in the case of the ion trap on. The discrepancy brought on by the collision energy is mitigated by the elastic ion-atom collision from the ions in the ion trap. $\gamma_r$ and $\gamma_x$ are linear, so the same conclusion is reached for $\gamma_r$.

\begin{figure}[htbp]
	\includegraphics[width=3.5in]{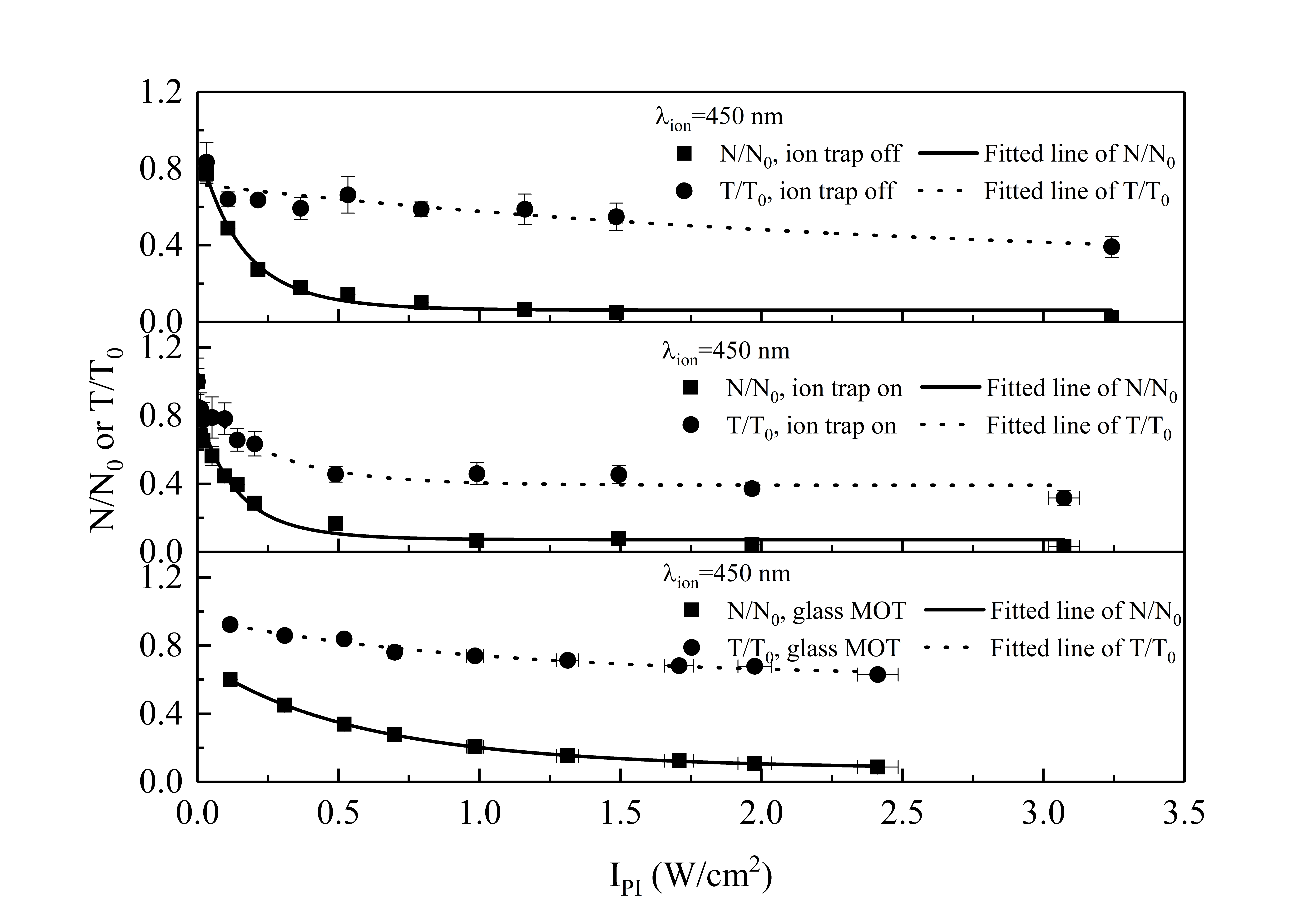}
	\caption{\label{N-T-I-450}The number and temperature of the remaining atoms as a function of the intensity of the ionization laser. Measurements are performed in the ion-neutral hybrid trap\cite{Lv2016cpl} with or without the ion trap, and in the standard vapor-loaded MOT with a glass vacuum rectangular chamber\cite{ruan-2014}, respectively. The wavelength of the ionization laser is 450 nm and the detuning $\Delta $ of the first excitation laser frequency from the transition $5~^2S_{1/2},F=2\rightarrow 5~^2P_{3/2},F^\prime=3$ is -12 MHz. The irradiation period of 4 s. $N_0$ and $T_0$ represent the number and the temperature of trapped atoms in a steady state without photoionization, respectively. The atomic number is fitted by a single exponential curve $ae^{-bI_{PI}}+c$, and the temperature is fitted by a double exponential curve $(ae^{-bI_{PI}}+c)^2$. The following graphs are so.}
\end{figure}

\begin{figure}[htbp]
	\includegraphics[width=3.5in]{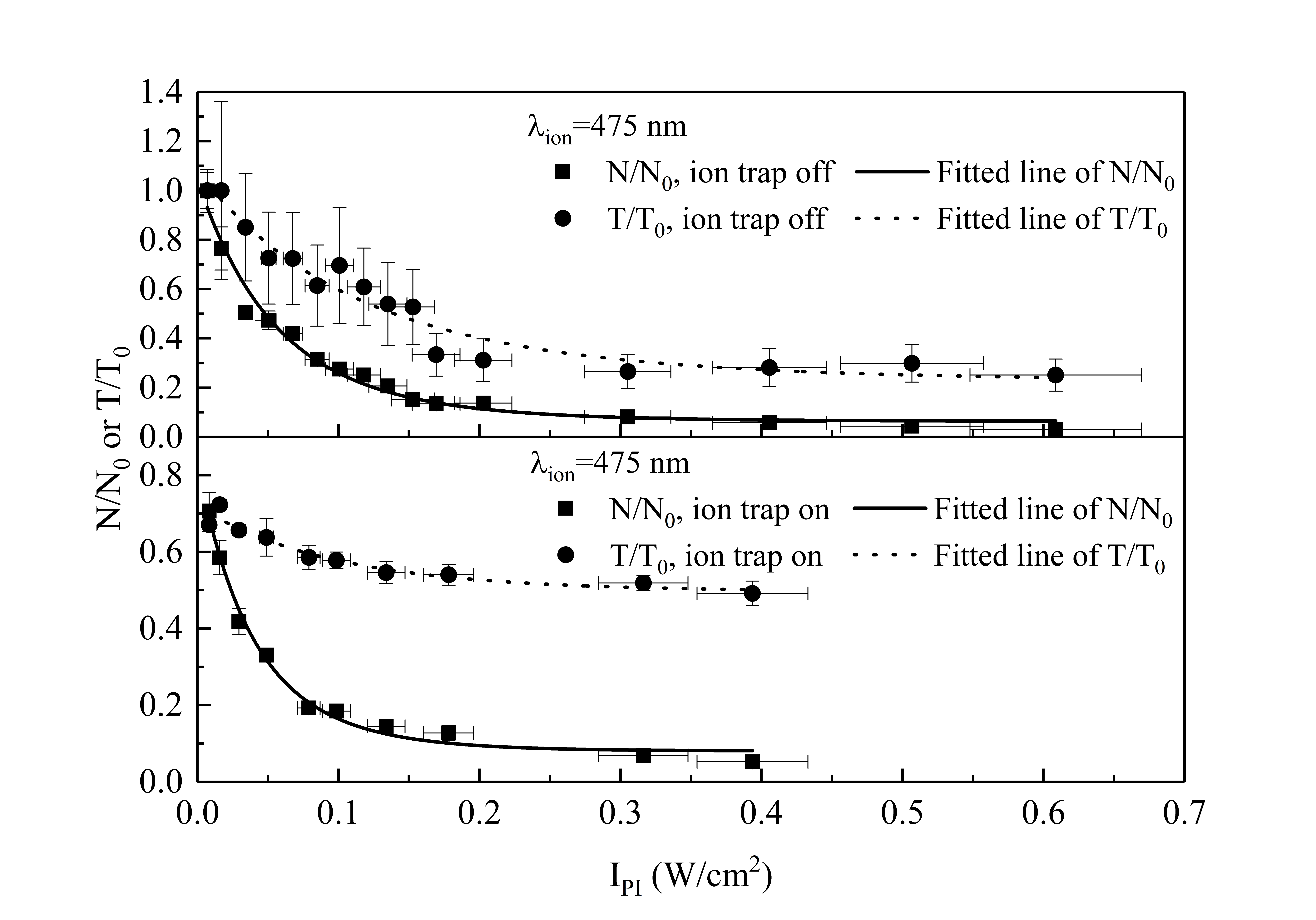}
	\caption{\label{N-T-I-475}The number of the remaining atoms of $N/N_{0}$ as a function of the intensity of the ionization laser in the two-step CW-laser photoionization of laser-cooled $^{87}$Rb atoms. Measurements are performed in ion-neutral hybrid trap \cite{Lv2016cpl} with or without the ion trap on. The wavelength of the ionization laser is 475 nm. The detuning $\Delta$ of the first excitation laser is -12 MHz. The irradiation period of the ionization laser is 4 s.}
\end{figure}

\begin{figure}[htbp]
	\includegraphics[width=3.5in]{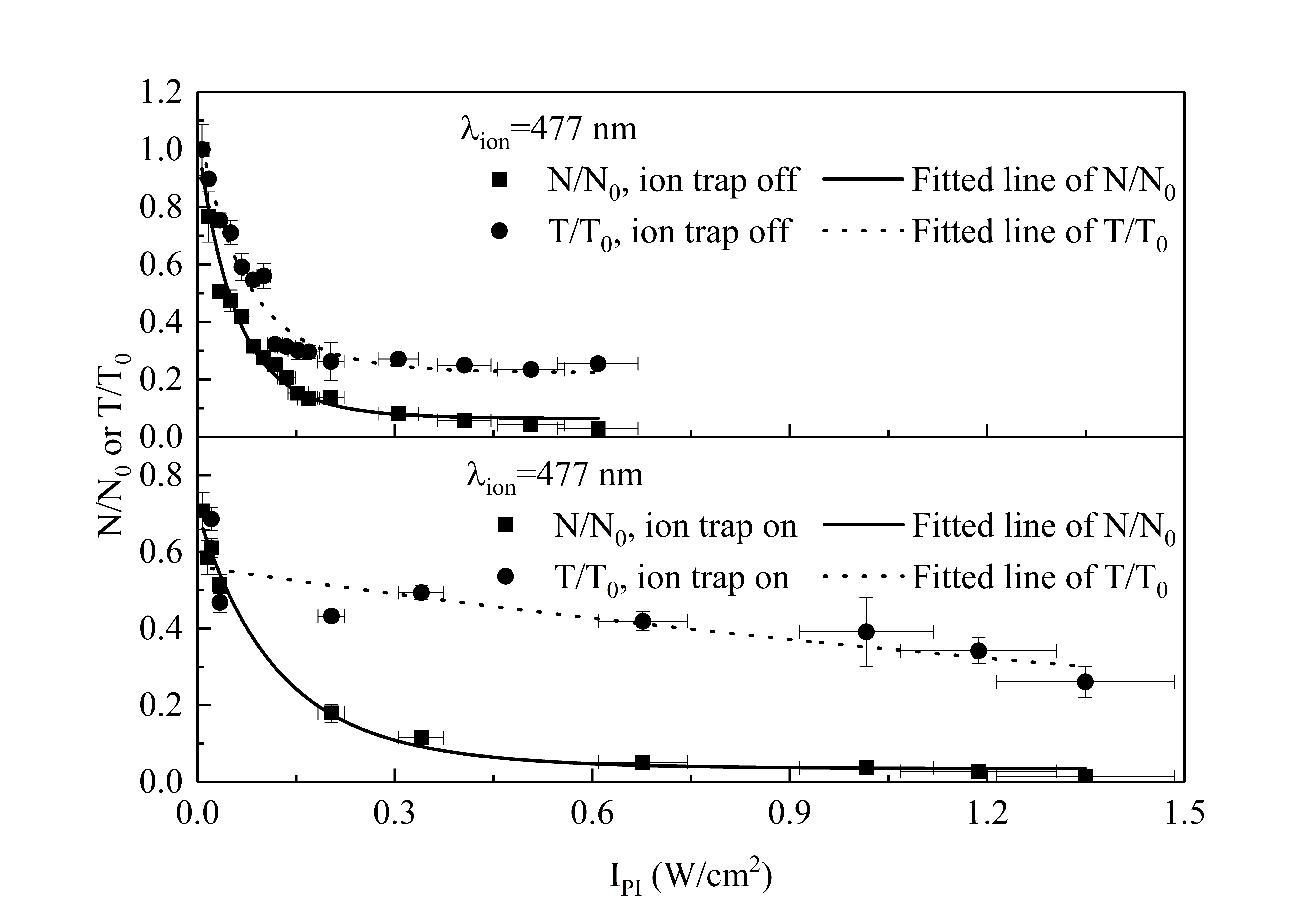}
	\caption{\label{N-T-I-477}The number of the remaining atoms of $N/N_{0}$ as a function of the intensity of the ionization laser in the two-step CW-laser photoionization of laser-cooled $^{87}$Rb atoms. Measurements are performed in ion-neutral hybrid trap \cite{Lv2016cpl} with or without the ion trap on. The wavelength of the ionization laser is 477 nm and the detuning $\Delta$ of the first excitation laser is -12 MHz. The irradiation period of the ionization laser is 4 s.}
\end{figure}

\begin{figure}[htbp]
	\includegraphics[width=3.5in]{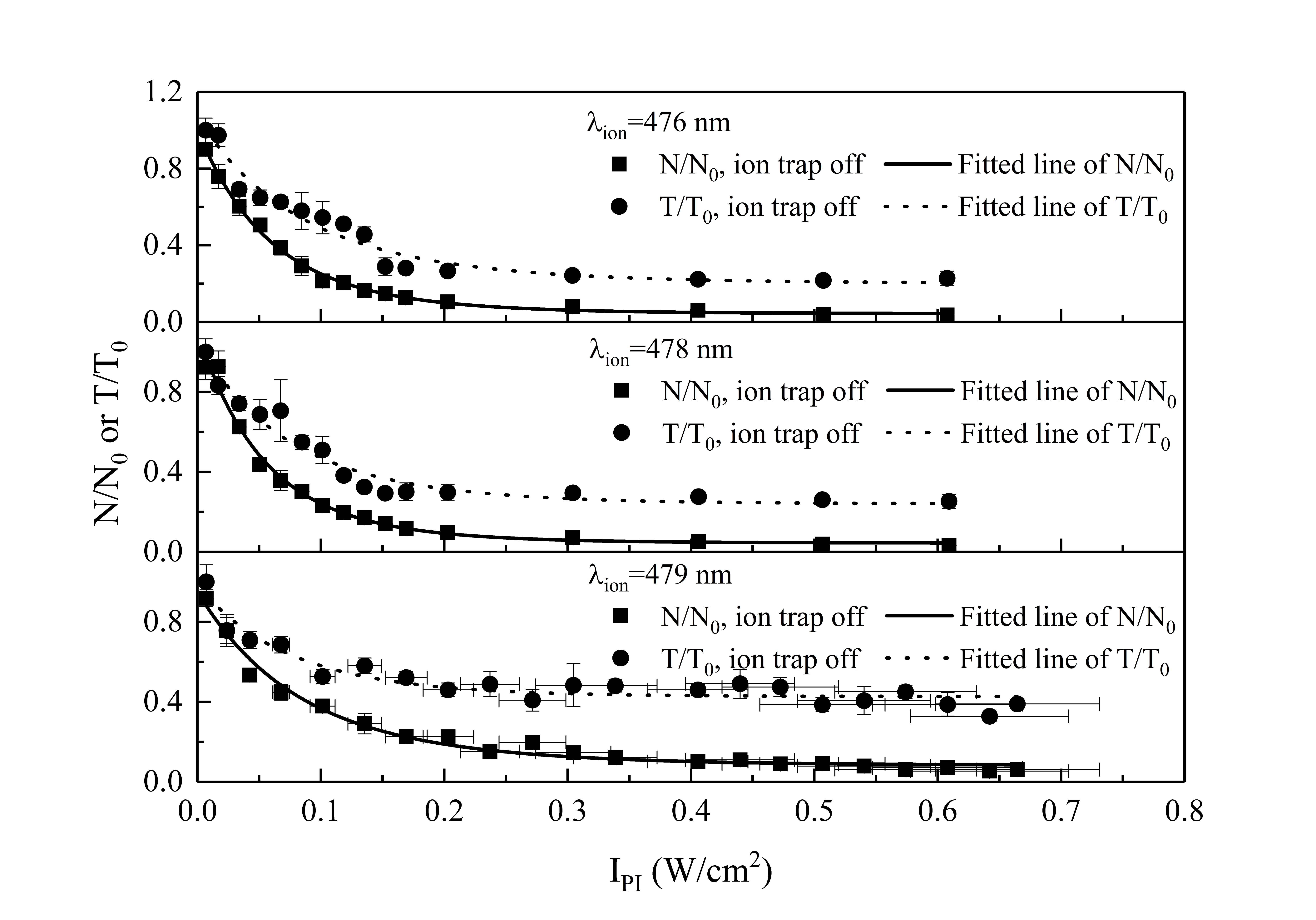}
	\caption{\label{N-T-I-476-478-479} The number of the remaining atoms of $N/N_{0}$ as a function of the intensity of the ionization laser in the two-step CW-laser photoionization of laser-cooled $^{87}$Rb atoms. Measurements are performed in ion-neutral hybrid trap \cite{Lv2016cpl} with or without the ion trap on. The wavelength of the ionization laser is 476(up), 478 (middle), and 479 (bottom) nm, respectively. The detuning $\Delta$ of the first excitation laser is -12 MHz. The irradiation period of the ionization laser is 4 s.}
\end{figure}

Now we survey the variation of atomic number and temperature with the intensity of the ionization laser in the case that the irradiation period of the ionization light was set as 4 s, the wavelength of the ionization laser was varied among 450, 475, 476, 477,478 and 479 nm, corresponding to initial electron temperatures of 1295.2, 171.4, 128.9, 86.6, 44.4, and 2.5 K, respectively. The ratios $N/N_0$ and $T/T_0$ are taken to overcome the fluctuation of MOT \cite{ruan-2014}. $N_0$ and $T_0$ represent the number and the temperature of trapped atoms in a steady state without photoionization, respectively. As demonstrated in Figs.\labelcref{N-T-I-450,N-T-I-475,N-T-I-477,N-T-I-476-478-479}, for any wavelength of the ionization laser, the variation of ratio in the atomic number $N/N_0$ is well fit by a single exponential decay function plus an offset, which can be explained by the Eq.\ref{eq:atomloss}. The variation of ratio in the temperature of atomic cloud $T/T_0$ is well fit by a double exponential decay function plus an offset because of $T\propto r_{atom}^2$ \cite{book1999}.

\begin{figure}[htbp]
	\includegraphics[width=3.5in]{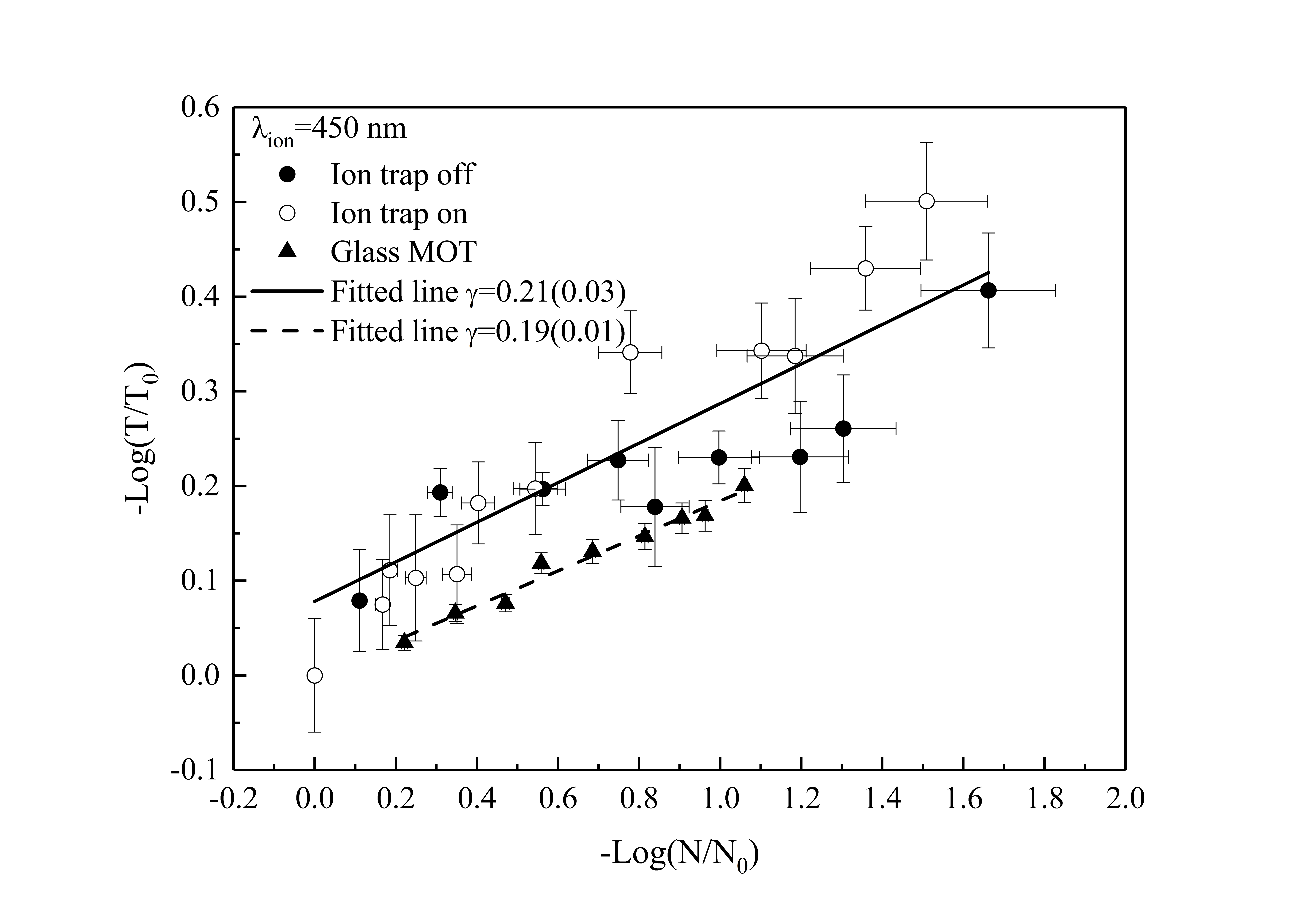}
	\caption{\label{N-T-450} The behavior of $-\log(T/T_{0})$ as a function of $-\log(N/N_{0})$ obtained with different $I_{PI}$ in the two-step CW-laser photoionization of laser-cooled $^{87}$Rb atoms with $\lambda_{ion}$=450 nm. Measurements are performed in ion-neutral hybrid trap \cite{Lv2016cpl} with or without the ion trap on and in the standard vapor-loaded MOT with a glass vacuum rectangular chamber\cite{ruan-2014}, respectively. The irradiation period of the ionization laser is 4 s. $T_0$ is the temperature of trapped atoms in a steady state, the following graphs are so.}
\end{figure}

\begin{figure}[htbp]
	\includegraphics[width=3.5in]{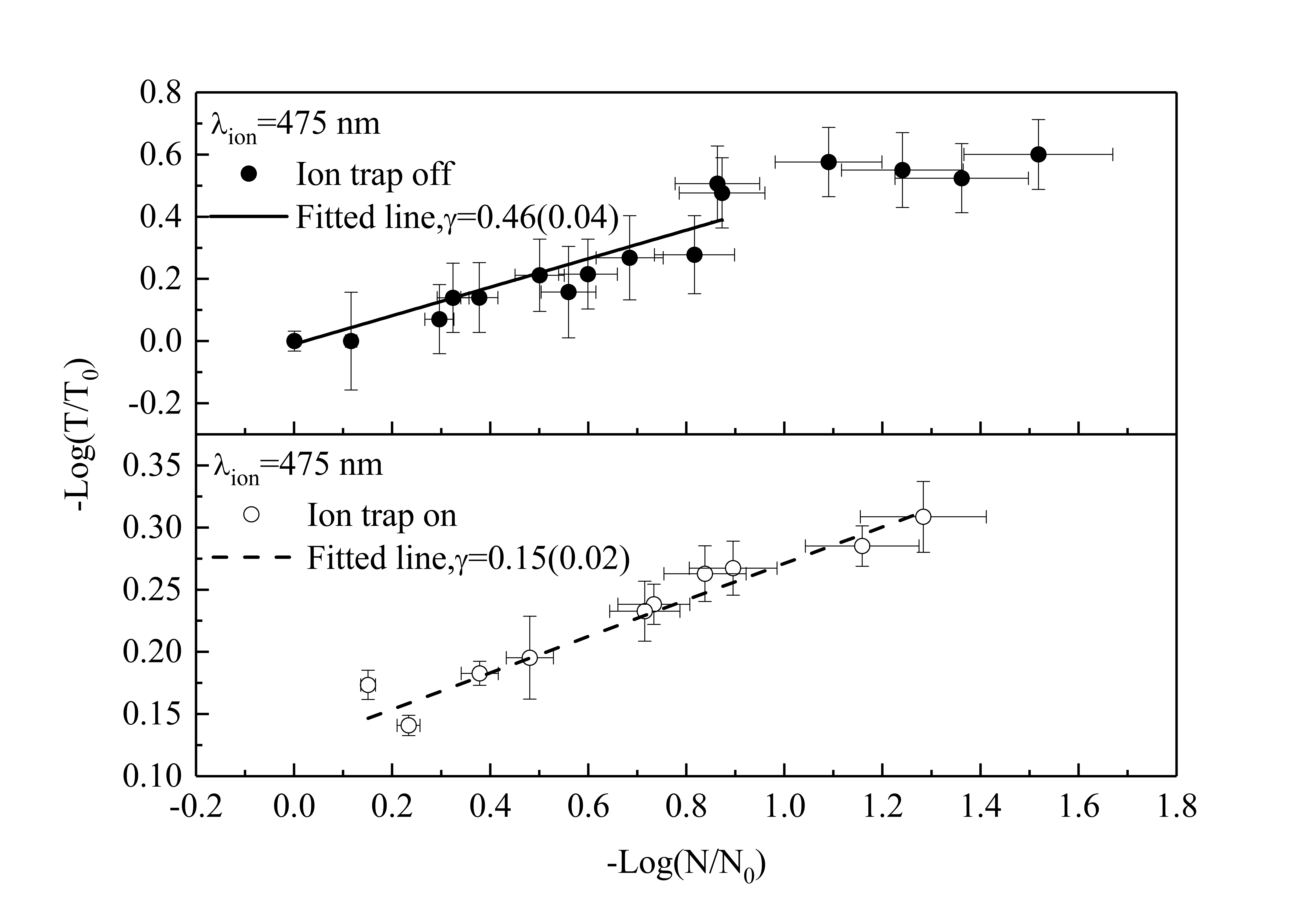}
	\caption{\label{N-T-475}The behavior of $-\log(T/T_{0})$ as a function of $-\log(N/N_{0})$ obtained with different $I_{PI}$ in the two-step CW-laser photoionization of laser-cooled $^{87}$Rb atoms with $\lambda_{ion}$=475 nm. Measurements are performed in ion-neutral hybrid trap \cite{Lv2016cpl} with or without the ion trap on. The wavelength of the ionization laser is 475 nm.  The irradiation period of the ionization laser is 4 s.}
\end{figure}

\begin{figure}[htbp]
	\includegraphics[width=3.5in]{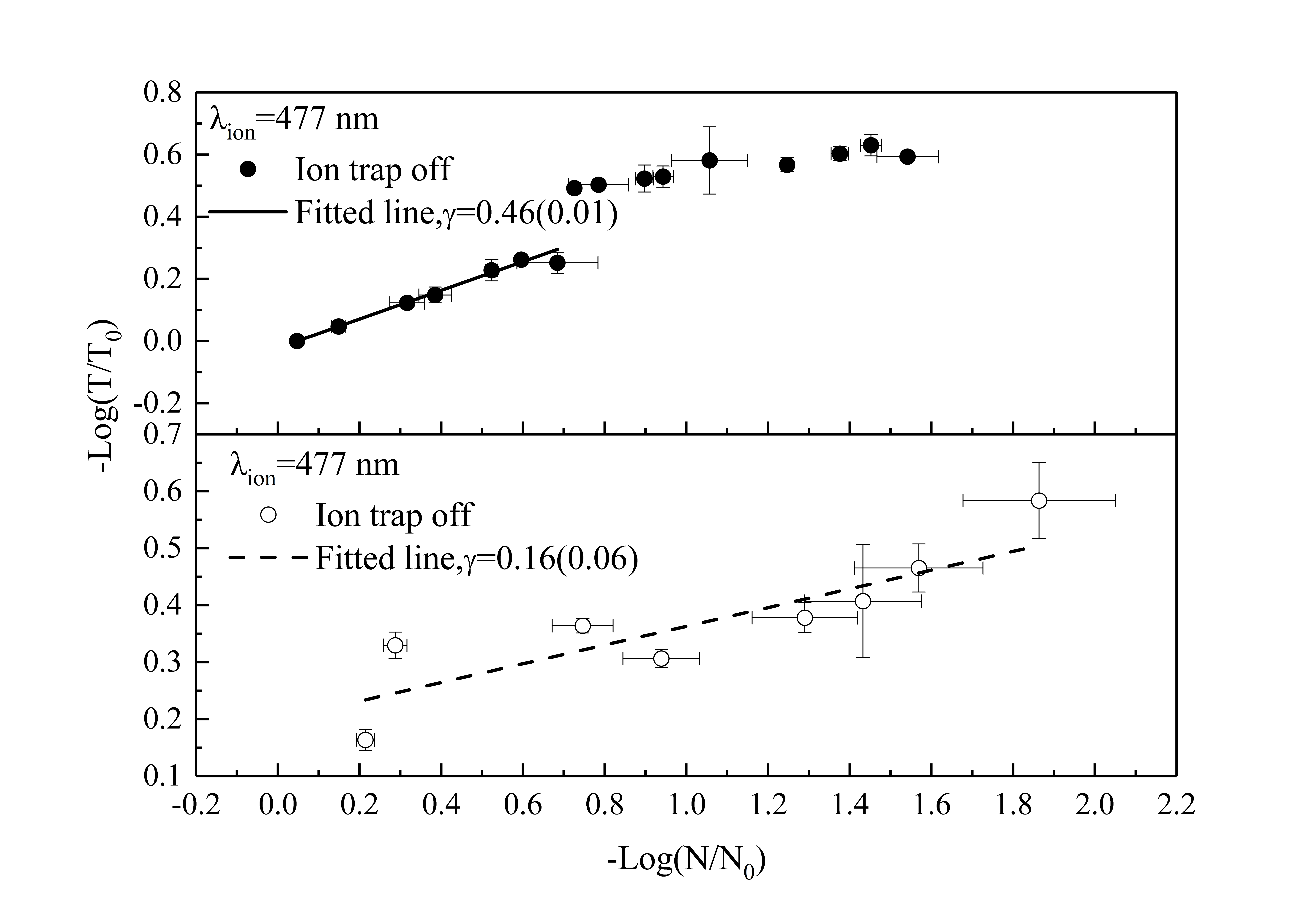}
	\caption{\label{N-T-477}The behavior of $-\log(T/T_{0})$ as a function of $-\log(N/N_{0})$ obtained with different $I_{PI}$ in the two-step CW-laser photoionization of laser-cooled $^{87}$Rb atoms with $\lambda_{ion}$=477 nm. Measurements are performed in ion-neutral hybrid trap \cite{Lv2016cpl} with or without the ion trap on. The wavelength of the ionization laser is 477 nm.  The irradiation period of the ionization laser is 4 s.}
\end{figure}

\begin{figure}[htbp]
	\includegraphics[width=3.5in]{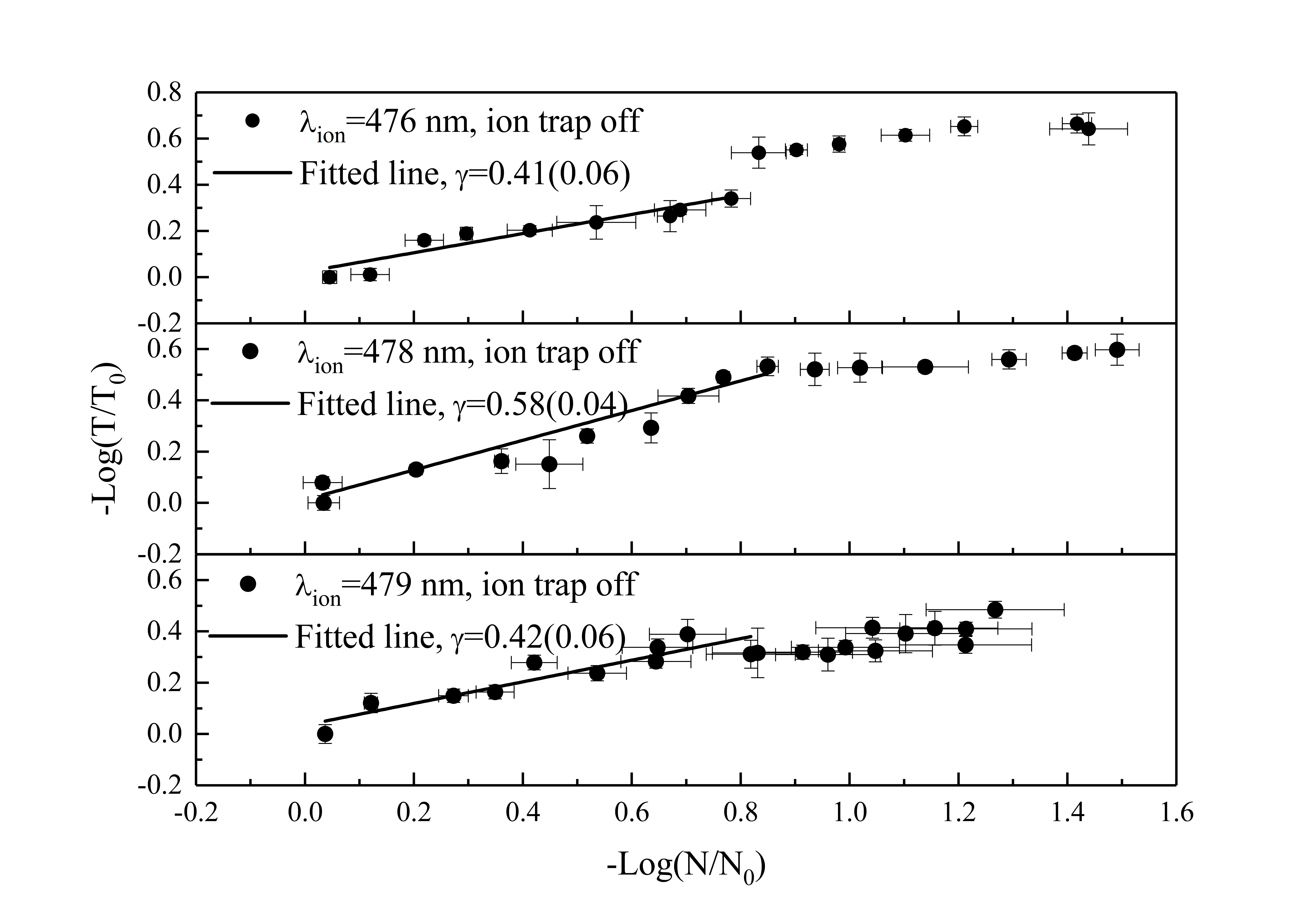}
	\caption{\label{N-T-476-478-479}The behavior of $-\log(T/T_{0})$ as a function of $-\log(N/N_{0})$ obtained with different $I_{PI}$ in the two-step CW-laser photoionization of laser-cooled $^{87}$Rb atoms with $\lambda_{ion}$=450 nm. Measurements are performed in the ion-neutral hybrid trap \cite{Lv2016cpl} without the ion trap on. The wavelength of the ionization laser is 476 (up), 478 (middle), and 479 (bottom) nm, respectively. The irradiation period of the ionization laser is 4 s.}
\end{figure}

As for the relationships between $-\log(T/T_0)$ and $-\log(N/N_0)$, it depends not only on the wavelength of the ionization laser but also on whether the ion trap is turned on as shown in Figs.\labelcref{N-T-450,N-T-475,N-T-477,N-T-476-478-479}. For the case that the wavelength of the ionization laser is 450 nm shown in Fig.\ref{N-T-450}, $-\log(T/T_0)$ linearly increases as $-\log(N/N_0)$, It means that the number $N$ and the temperature $T$ of trapped atoms follows a power law $T\propto N^\gamma$. More interestingly, the data obtained with or without the ion trap on almost fall on the same curve within the experimental error. The $\gamma$ obtained in the ion-neutral hybrid trap is 0.21(0.03) and this is nearly the same as that 0.19(0.01) obtained in the glass MOT\cite{ruan-2014}. The difference between this work measured in the hybrid trap and the result measured in the MOT with a glass chamber\cite{ruan-2014} is the difference in the offset. This may be mainly due to different collisional loss rates $\gamma_L$ for the two apparatuses.

\begin{figure}[htbp]
	\includegraphics[width=3.5in]{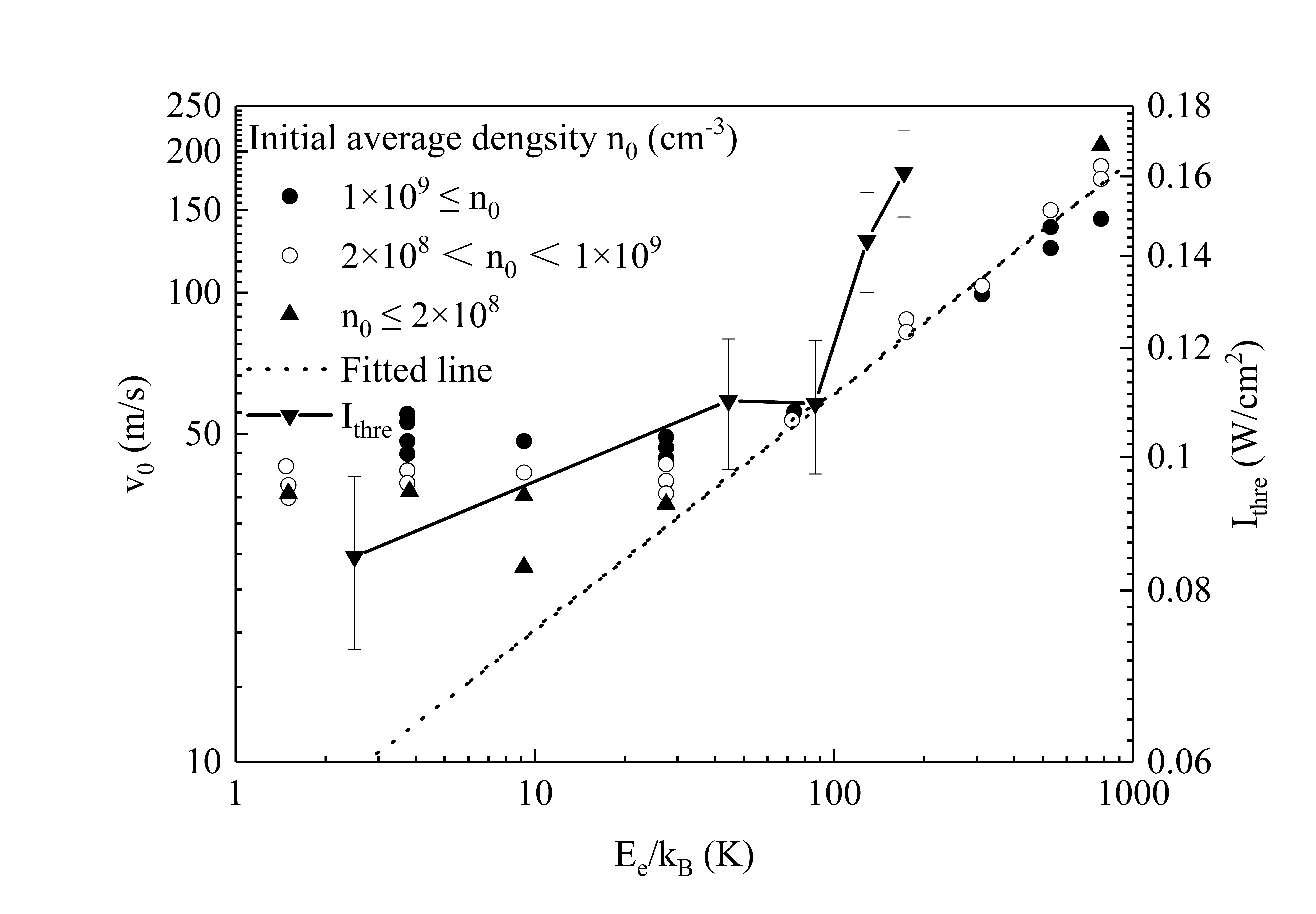}
	\caption{\label{threshold}Comparsion of expansion velocities of ions $v_0$\cite{ion-v-2000} and the ionization laser intensity at the onset $I_{thre}$ as a function of initial electron temperature.}
\end{figure}

The relationships between $-\log(T/T_0)$ and $-\log(N/N_0)$ are similar when the wavelength of the ionization laser was varied among 475, 476, 477, 478 and 479 nm, as shown in Figs.\labelcref{N-T-475,N-T-477,N-T-476-478-479}. We take the 477 nm ionization laser case as an example and show it in Fig.\ref{N-T-477}. In the case that the ion trap is turned off, at first, $-\log(T/T_0)$ linearly increases as $-\log(N/N_0)$ increasing, then jumps at a certain ionization intensity $I_{thre}$ and remains a constant after the jump. However, when the ion trap is turned on, $-\log(T/T_0)$ linearly increases as $-\log(N/N_0)$ increasing in the whole range of experimental ionization laser intensity. To reveal the underlying physical mechanism, we compare the expansion velocities of ions in an ultracold neutral plasma as a function of the initial electron temperature\cite{ion-v-2000} with the intensity of the ionization laser at the onset of curves for $-\log(T/T_0)$ versus $-\log(N/N_0)$, as shown in Fig.\ref{threshold}, and their variation behavior is similar. This result further indicates that electrons affect ion-atom collisions through disorder-induced heating in CW-laser photoionization. Specifically, the linear relationship between $-\log(T/T_0)$ and $-\log(N/N_0)$ is the result from the fact that $N$ and $T$ follows a power law $T\propto N^\gamma$. Jumping to a constant value means that temperature achieves equilibrium when the irradiation period of the ionization laser is 4 s. It illustrates a jump in the behavior of decreasing rates of the temperature $\gamma_r$. By taking DIH into account, it is clearly understood. As was already mentioned, the rate of atomic loss $\gamma_x$ should increase with the intensity of the ionization laser, the DIH effect intensifies, and the increase in $\gamma_r$ follows. Thus, the temperature approaches equilibrium more quickly as the intensity of the ionization laser increases.

\section{Conclusion}

To investigate the impact of ion-atom elastic collisions on the number and temperature of the remaining atoms in CW-laser photoionization, we measured the number of the remaining atoms, radius, and temperature of the cold atomic cloud in an Rb$^+$-Rb hybrid trap. We have investigated the dependence of the number and temperature of the remaining atoms on the wavelength and intensity of the ionization laser as well as the ion trap on or off. We discover that there is good agreement between fits with a single exponential decay function plus an offset to the number and radius of remaining atoms, respectively. The two exponential factors $\gamma_x$ and $\gamma_r$ follow a well-linear connection since the temperature $T$ and number $N$ of remaining atoms in an MOT follow a power law, $T\propto N^\gamma$. Therefore, the evolution of the number and temperature of the remaining atoms can be obtained by analyzing the loss rate of atoms. The ion-atom elastic collision will cause atomic loss. We believe that electrons can affect ion-atom collisions by disorder-induced heating (DIH) caused by spatial correlations between electrons and ions. Specifically, even if it cannot produce plasma, DIH still heats the ions. The ion temperature determines the ion-atom collision energy and thus affects the number and temperature of remaining atoms. On the other hand, when the ion trap turns on to repel the electrons or the initial electron temperature is above 1000 K, there is no DIH. These results are shown in Figs.\labelcref{N-T-1A,N-T-1B,N-T-450,N-T-475,N-T-477,N-T-476-478-479} and reveal that both the effect of ion-atom collisions and the existence of electrons on the temperature $T$ and number $N$ of the remaining atoms. These results demonstrate that we created a powerful experimental tool, i.e., the number and temperature of the remaining atoms, to study the relaxation of ion temperature and exhibit the effect of ion-atom elastic collisions on the number and temperature of the remaining atoms in CW-laser photoionization. Our finding opens new avenues for research into atomic processes, transport, etc. in ultracold plasma. This has an important reference for high-energy-density plasma (HEDP) research.

\clearpage

\section{Acknowledgements}

This study was supported by the National Key Research and Development Program of China (Grant Nos. 2017YFA0402300 and 2017YFA0304900), the Beijing Natural Science Foundation (Grant No. 1212014), the Fundamental Research Funds for the Central Universities, the Key Research Program of the Chinese Academy of Sciences (Grant No. XDPB08-3), specialized research fund for CAS Key Laboratory of Geospace Environment (GE2020-01), and National Natural Science Foundation of China (61975091, 61575108).

\clearpage
\bibliographystyle{unsrt}
\bibliography{N-V.bib}
\end{document}